\documentclass[twocolumn]{aastex701}
\usepackage{amsmath,amssymb,amsfonts}
\usepackage{booktabs}
\usepackage{microtype}
\usepackage{graphicx}
\usepackage{url}
\usepackage{xcolor}

\shorttitle{Asteroseismic Detection of a Massive Unseen Companion}
\shortauthors{Lv et al.}

\begin{document}

\title{\textbf{Asteroseismic Detection of a Massive Unseen Companion to the $\delta$ Scuti Star TIC~160582982}}

\author[orcid=0000-0001-6354-1646]{Chenglong Lv}
\affiliation{Xinjiang Astronomical Observatory, Chinese Academy of Sciences, Urumqi, Xinjiang 830011, People's Republic of China}
\affiliation{Instituto de Astrof\'isica de Andaluc\'ia - CSIC, 18008 Granada, Spain}
\email[show]{lvchenglong@xao.ac.cn}

\author[orcid=0000-0003-0139-6951]{J. Pascual-Granado}
\affiliation{Instituto de Astrof\'isica de Andaluc\'ia - CSIC, 18008 Granada, Spain}
\email{j.pascual@csic.es}

\author[orcid=0000-0003-1845-4900]{Ali Esamdin}
\affiliation{Xinjiang Astronomical Observatory, Chinese Academy of Sciences, Urumqi, Xinjiang 830011, People's Republic of China}
\affiliation{School of Astronomy and Space Science, University of Chinese Academy of Sciences, Beijing 100049, People's Republic of China}
\email{aliyi@xao.ac.cn}

\author[orcid=0000-0002-3234-9130]{Angela Gardini}
\affiliation{Instituto de Astrof\'isica de Andaluc\'ia - CSIC, 18008 Granada, Spain}
\email{gardini@iaa.es}

\author[orcid=0000-0002-2463-4943]{Haozhi Wang}
\affiliation{Chinese Academy of Sciences South America Center for Astronomy}
\email{wanghaozhi@xao.ac.cn}

\author[orcid=0009-0000-2772-2677]{Jianwei Yao}
\affiliation{Xinjiang Astronomical Observatory, Chinese Academy of Sciences, Urumqi, Xinjiang 830011, People's Republic of China}
\affiliation{School of Astronomy and Space Science, University of Chinese Academy of Sciences, Beijing 100049, People's Republic of China}
\email{yaojianwei@xao.ac.cn}

\author[orcid=0000-0001-5066-5682]{Shuguo Ma}
\affiliation{Xinjiang Astronomical Observatory, Chinese Academy of Sciences, Urumqi, Xinjiang 830011, People's Republic of China}
\email{mashuguo@xao.ac.cn}

\begin{abstract}
We present TIC~160582982, an eccentric $\delta$ Scuti binary with a dynamically massive unseen companion. Six coherent high frequency pressure modes used as pulsation clocks yield a phase modulation orbit with $P_{\rm orb}=89.29\pm0.19~{\rm d}$, $e=0.539_{-0.037}^{+0.040}$, and $a_1\sin i/c=145.6_{-3.4}^{+3.7}~{\rm s}$, corresponding to $K_1=42.2_{-1.8}^{+2.2}~{\rm km~s^{-1}}$ and $f(M)=0.416_{-0.028}^{+0.032}~M_\odot$. An independent frequency-modulation analysis of the two strongest modes gives a consistent orbital solution. Rotating MIST isochrones constrained by the atmospheric parameters and spectral energy distribution (SED) give $M_1=1.93\pm0.19~M_\odot$ and $R_1=2.65\pm0.23~R_\odot$. The mass function implies a formal edge-on minimum companion mass of approximately $1.80~M_\odot$. Assuming a luminous main-sequence companion and including the non-eclipsing geometry gives $M_2=2.19_{-0.34}^{+1.24}~M_\odot$ and $i=61.5_{-19.6}^{+16.6}~{\rm deg}$. However, no clear evidence for the presence of a luminous counterpart is found, either photometrically or spectroscopically. This suggests the presence of a massive compact object, such as a neutron star or even a stellar-mass black hole, as the massive unseen binary companion. Future phase resolved high resolution spectroscopy will be crucial for obtaining an independent orbital constraint and clarifying the nature of the companion.
\end{abstract}


\section{Introduction}
\label{sec:intro}

A growing number of dormant compact object candidates have been identified
through the motion of a luminous companion rather than through accretion.
This approach is essential for detached systems, where X-ray or radio
emission can be weak or absent. Astrometric and spectroscopic discoveries
such as Gaia BH1 and Gaia BH2 have shown that noninteracting black holes can
be found in wide binaries, but they also highlight the difficulty of
confirming an unseen companion when only one luminous star is directly
observed
\citep{ElBadry2023a,ElBadry2023b,Chakrabarti2023,Tanikawa2023}.
The discovery of Gaia BH3 further demonstrates the power of astrometric
searches for dormant black holes in detached binaries
\citep{GaiaCollaboration2024BH3}. At lower companion masses, the problem is
even more delicate. The interval between the most massive neutron stars and
the lightest black holes remains observationally sparse and theoretically
uncertain
\citep{Ozel2010,Farr2011,Kreidberg2012,Abbott2020}.
Recent electromagnetic searches have also identified wide binaries with
dynamically massive and apparently faint companions, showing that this mass
range can be investigated outside accreting X-ray binaries and
gravitational wave events \citep{Wang2024MassGapBH}. Asteroseismic searches for quiescent compact companions using pulsation phase or frequency modulation have also been discussed by \citet{ShibahashiMurphy2020QuietBH}.

Pulsating stars offer an independent way to measure binary motion. If a
pulsation mode is coherent, the changing light travel time across the orbit
appears as a phase modulation of the oscillation. This method turns the
pulsation into a clock: the observed time delay gives the projected
semimajor axis of the pulsating star and hence the binary mass function
\citep{ShibahashiKurtz2012,Murphy2014,Murphy2016}. The recovery of binary orbital properties from pulsation light-arrival-time delays was further developed by \citet{MurphyShibahashi2015PM}. The rich pressure mode
spectra of many $\delta$ Scuti stars make them useful probes of stellar
structure and binarity \citep[e.g.,][]{Aerts2010,Guzik2021}. Pulsation timing
has been applied extensively to \textit{Kepler} $\delta$ Scuti stars
\citep{Murphy2018,Murphy2020} and more recently to the Transiting Exoplanet
Survey Satellite (\textit{TESS}) archive
\citep{Ricker2015,Dholakia2025}. For systems that are not eclipsing and have
limited spectroscopy, phase modulation can provide the first dynamical
evidence for an unseen companion.

TIC~160582982 is an especially interesting target for this approach. It is a
rapidly rotating A-type $\delta$ Scuti star with coherent high frequency
pulsations and a long, eccentric phase modulation orbit \citep{Dholakia2025}. The inferred mass
function requires a massive dynamical companion, whereas the available
optical to mid infrared SED is adequately described by a single A-type
photosphere and the archival TRES classification products show no obvious
secondary component. This combination motivates a quantitative assessment
of the companion light, although neither the broadband SED nor the existing
spectroscopic products alone provide a stringent exclusion of a luminous
secondary.

In this work, we derive the orbit from the phase modulation (PM) of six high frequency pressure modes and independently test the orbital solution using frequency modulation (FM) of the two strongest modes. We then combine the timing constraints with rotating MIST constraints on the visible primary and use the broadband SED, a first order companion light calculation, and archival TRES diagnostics to evaluate the allowed companion configurations, including a single luminous star, an unresolved inner binary, and a compact remnant such as a neutron star or stellar-mass black hole. Finally, we examine the orbital phase dependence of a low frequency signal as a possible indication that the eccentric orbit influences the pulsations of the visible star.

\section{Observations and Data Analysis}
\label{sec:data}
\setcounter{footnote}{0}

We used the 2-min cadence TESS light curves produced by the Science Processing Operations Center (SPOC; \citealt{Jenkins2016}) in Sectors 14--20 and 22--26. The 12 light curve segments span $\sim$ 352~d, and we adopted $\mathrm{BJD}_{\mathrm{TDB}}=2458683.347307$ as the reference epoch. The downloaded SPOC light curve files contain 222,761 cadences in total. After applying the data quality selection, 194,661 measurements were retained for the frequency analysis, corresponding to 87.4\% of the downloaded cadences. Relative to uninterrupted 2-min sampling over the full time baseline, the effective duty cycle is approximately 76.9\%.

For a 2-min cadence, the nominal Nyquist frequency is \textbf{$\nu_{\mathrm{Nyq}}=360~\mathrm{d^{-1}}$}, well above the frequency range relevant to the present analysis. We restricted the pulsation frequency extraction to \textbf{$0<\nu<80~\mathrm{d^{-1}}$}, which encompasses the observed low frequency variability and the pressure mode spectrum of the $\delta$~Scuti primary. We calculated a Lomb--Scargle amplitude spectrum with \textsc{FELIX} \citep{Charpinet2010}. Frequencies were extracted through iterative pre-whitening. At each iteration, the highest remaining significant peak was identified, and the light curve was fitted with the multi sinusoidal model
\begin{equation}
x(t)=x_0+\sum_{i=1}^{N}A_i
\sin\!\left[2\pi\left({\nu_i} t+\phi_i\right)\right],
\label{eq:multisine}
\end{equation}
where $x_0$ is a constant flux offset and $A_i$, \textbf{$\nu_i$}, and $\phi_i$ are, respectively, the amplitude, frequency, and phase of the $i$th component. The fitted signal was subtracted from the light curve, and the residual amplitude spectrum was recalculated before the next iteration. We adopted a detection threshold of $\mathrm{S/N}>5.2$, following the recommendations for space based photometric time series \citep{Baran2015,Baran2021}. The extraction was terminated when no additional peak exceeded this threshold, yielding 128 significant frequencies. Formal uncertainties in the fitted frequencies, amplitudes, and phases were estimated using the analytic least squares expressions of \citet{Montgomery1999}. The complete frequency list and fitted parameters are provided in Appendix~\ref{app:frequency_list}.

Several high-amplitude pressure modes are accompanied by nearly symmetric sidelobes separated by approximately the orbital frequency. Table~\ref{tab:orbital_multiplets} summarizes the six central peaks used for the PM analysis and the two clearest quintuplets used for the FM analysis. The latter are the multiplet structures expected from orbital frequency modulation \citep{Shibahashi2015FM}.

\begin{deluxetable*}{lrrrr}
\tabletypesize{\footnotesize}
\tablecaption{Pulsation Multiplets at Family-specific Phase-zero Epochs\label{tab:orbital_multiplets}}
\tablehead{
\colhead{Component} &
\colhead{Frequency} &
\colhead{Amplitude} &
\colhead{Phase} &
\colhead{$\Delta\phi$} \\
\colhead{} &
\colhead{($\mathrm{d}^{-1}$)} &
\colhead{(mmag)} &
\colhead{(rad)} &
\colhead{(rad)}
}
\startdata
\sidehead{\hfill $\nu_1$: $T_0=2458688.05574$}
\hline \\[-2ex]
$\nu_1-2\nu_{\rm orb}$ & 36.01567(23) & 0.0119(17) & $2.027\pm0.289$ &  \\
$\nu_1-\nu_{\rm orb}$ & 36.02687(5) & 0.0536(17) & $1.410\pm0.064$ &  \\
$\nu_1$ & 36.038013(9) & 0.2963(17) & $-0.117\pm0.012$ & $1.527\pm0.049$ \\
$\nu_1+\nu_{\rm orb}$ & 36.04917(6) & 0.0493(17) & $1.410\pm0.070$ & $0.000$ \\
$\nu_1+2\nu_{\rm orb}$ & 36.06041(18) & 0.0150(17) & $1.132\pm0.232$ & $-0.896\pm0.370$ \\
\cutinhead{$\nu_2$: $T_0=2458701.42381$}
$\nu_2-2\nu_{\rm orb}$ & 34.26939(28) & 0.0095(17) & $0.812\pm0.357$ &  \\
$\nu_2-\nu_{\rm orb}$ & 34.28083(6) & 0.0453(17) & $-1.564\pm0.075$ &  \\
$\nu_2$ & 34.292075(10) & 0.2616(17) & $3.037\pm0.013$ & $1.682\pm0.057$ \\
$\nu_2+\nu_{\rm orb}$ & 34.30331(6) & 0.0406(17) & $-1.564\pm0.083$ & $0.000$ \\
$\nu_2+2\nu_{\rm orb}$ & 34.31445(26) & 0.0102(17) & $0.209\pm0.329$ & $-0.603\pm0.486$ \\
\cutinhead{$\nu_3$: $T_0=2458694.54055$}
$\nu_3-2\nu_{\rm orb}$ & 24.11905(28) & 0.0096(17) & $-1.598\pm0.357$ &  \\
$\nu_3-\nu_{\rm orb}$ & 24.13050(9) & 0.0311(17) & $1.652\pm0.110$ &  \\
$\nu_3$ & 24.141682(10) & 0.2564(17) & $0.197\pm0.013$ & $1.455\pm0.083$ \\
$\nu_3+\nu_{\rm orb}$ & 24.15289(9) & 0.0281(17) & $1.652\pm0.121$ & $0.000$ \\
\cutinhead{$\nu_4$: $T_0=2458686.55010$}
$\nu_4-\nu_{\rm orb}$ & 39.86583(16) & 0.0174(18) & $-0.446\pm0.205$ &  \\
$\nu_4$ & 39.877242(22) & 0.1251(18) & $0.956\pm0.029$ & $-1.403\pm0.137$ \\
$\nu_4+\nu_{\rm orb}$ & 39.88865(14) & 0.0204(18) & $-0.446\pm0.175$ & $0.000$ \\
\cutinhead{$\nu_5$: $T_0=2458664.09195$}
$\nu_5-2\nu_{\rm orb}$ & 43.17605(30) & 0.0102(19) & $-3.045\pm0.379$ &  \\
$\nu_5-\nu_{\rm orb}$ & 43.18720(13) & 0.0230(19) & $1.492\pm0.167$ &  \\
$\nu_5$ & 43.198549(26) & 0.1128(19) & $2.098\pm0.034$ & $-0.607\pm0.128$ \\
$\nu_5+\nu_{\rm orb}$ & 43.20976(14) & 0.0209(19) & $1.492\pm0.181$ & $0.000$ \\
\cutinhead{$\nu_6$: $T_0=2458688.58928$}
$\nu_6-\nu_{\rm orb}$ & 32.09647(16) & 0.0171(17) & $3.094\pm0.199$ &  \\
$\nu_6$ & 32.108005(24) & 0.1105(17) & $-2.816\pm0.031$ & $-0.373\pm0.138$ \\
$\nu_6+\nu_{\rm orb}$ & 32.11933(14) & 0.0188(17) & $3.094\pm0.181$ & $0.000$ \\
\enddata
\tablecomments{
Parenthesized uncertainties on frequency and amplitude are formal $1\sigma$
errors in the last quoted digits. The family-specific epochs $T_0$ are given
in ${\rm BJD}_{\rm TDB}$. Phases are in radians at the listed family-specific
nominal $T_0$. For a central-mode row,
$\Delta\phi=\langle\phi_{\pm1}\rangle-\phi_0$; for a positive-sidelobe row,
$\Delta\phi=\phi_{+m}-\phi_{-m}$. The equality of the $m=\pm1$ phases is
imposed by the epoch choice and is not independent FM evidence. Phase errors
are conditional on fixed $T_0$ because the full covariance is unavailable.
Only $\nu_1$ and $\nu_2$ are used for the FM orbital solution, which adopts
a common epoch.
}
\end{deluxetable*}

We constructed an optical to mid infrared SED using archival broadband photometry from Tycho and ground based Johnson $B,V$ measurements, Gaia DR2 $G_{\rm BP}$, $G$, and $G_{\rm RP}$ photometry, the TESS bandpass, the Two Micron All Sky Survey (2MASS; \citealt{Skrutskie2006}) $JHK_s$ bands, and the Wide-field Infrared Survey Explorer (WISE; \citealt{Wright2010}) W1 and W2 bands. Together with the Gaia parallax, these data constrain the effective temperature, radius, and extinction of the visible star, and provide a test for additional optical or infrared continuum light from a possible luminous companion.

We also inspected the two archival TRES spectra available through ExoFOP-TESS for TIC~160582982. TRES is mounted on the 1.5 m Tillinghast reflector at the Fred Lawrence Whipple Observatory (FLWO; \citealt{SzentgyorgyiFuresz2007,Mink2011}) and provides high resolution optical spectroscopy at $R\simeq44{,}000$. We use the associated classification diagnostics only as qualitative context for the visible primary and to assess whether obvious secondary spectral features are present.

\section{Pulsation-Timing Orbit}
\label{sec:pm_orbit}

\begin{figure*}[t]
\centering
\includegraphics[width=0.95\textwidth]{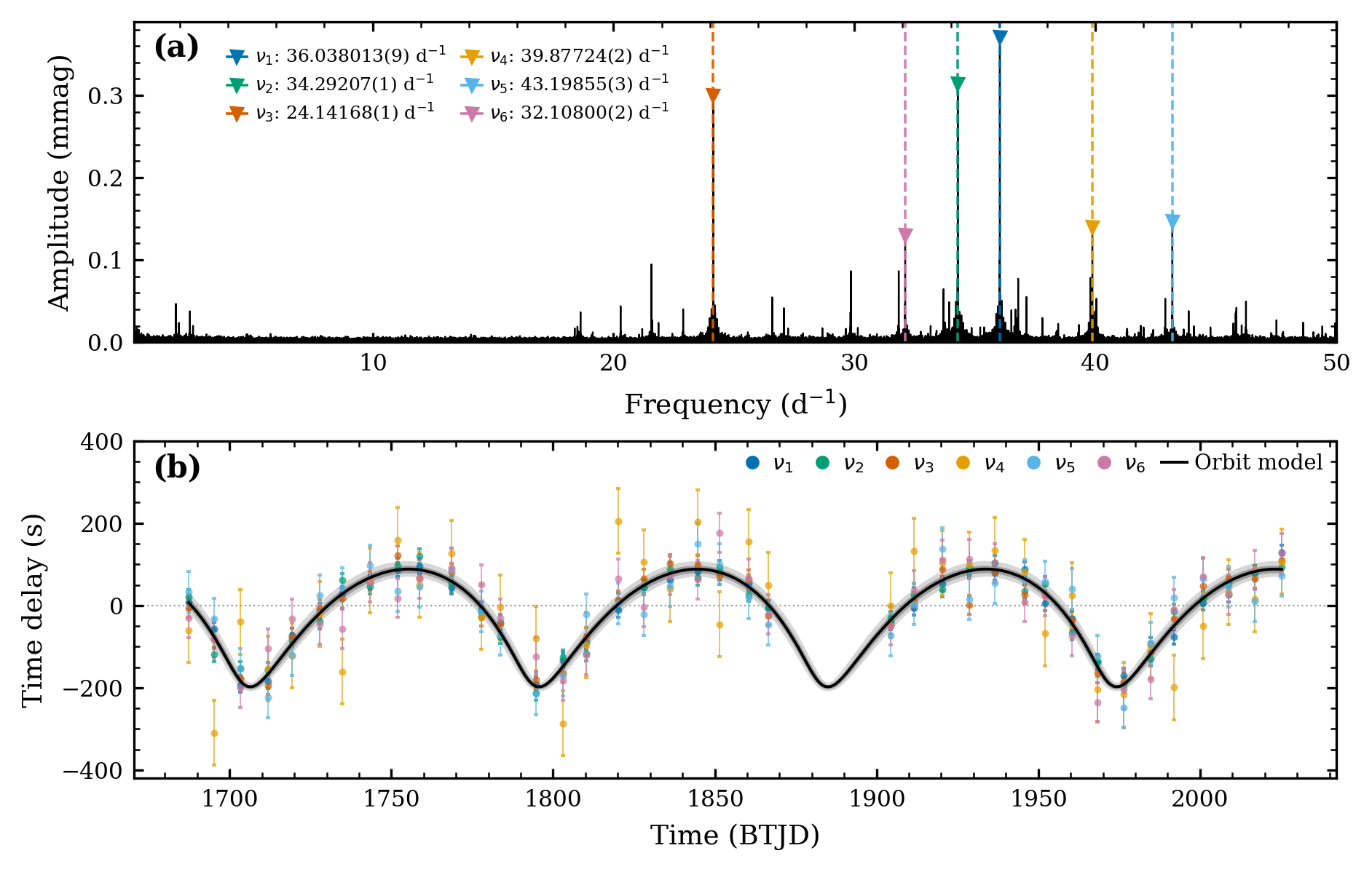}
\caption{Phase modulation orbit of TIC~160582982.
(a) Fourier amplitude spectrum of the TESS light curve. The six pressure modes used for the time delay extraction are marked.}
(b) Time delay measurements from the six selected modes. The black curve shows the best fitting eccentric light time orbit, with shaded regions indicating the model credible intervals.
\label{fig:orbital_pm}
\end{figure*}

We used the high frequency $\delta$ Scuti pulsations as coherent clocks to measure the orbital light travel time signal \citep{ShibahashiKurtz2012,Murphy2014,MurphyShibahashi2015PM, Murphy2016}. The Fourier amplitude spectrum contains several well resolved pressure modes. We selected six pressure modes and analyzed them with \texttt{Maelstrom} \citep{Hey2020}. For each mode, the phase variation was converted to a time delay using $\tau=\Delta\phi/(2\pi\nu)$. We then fitted the six time delay series simultaneously with a common Keplerian light time orbit while allowing mode specific phase offsets. The resulting time delays trace the line-of-sight displacement of the pulsating primary.

Figure~\ref{fig:orbital_pm} summarizes the phase modulation solution. The six selected modes give mutually consistent time delay curves, supporting an orbital origin rather than mode specific phase variability. The best fitting eccentric light time orbit has \textbf{$P_{\rm orb}=89.29\pm0.19~{\rm d}$}, \textbf{$e=0.539_{-0.037}^{+0.040}$}, and \textbf{$a_1\sin i/c=145.6_{-3.4}^{+3.7}~{\rm s}$}. These parameters correspond to \textbf{$a_1\sin i=0.2918_{-0.0067}^{+0.0075}~{\rm au}$}, a timing inferred radial velocity semiamplitude of \textbf{$K_1=42.2_{-1.8}^{+2.2}~{\rm km~s^{-1}}$}, and a binary mass function of \textbf{$f(M)=0.416_{-0.028}^{+0.032}~M_\odot$}. This mass function is large for an A-type primary and requires a massive companion, with still larger companion masses implied at lower inclinations.

\begin{figure*}[t]
\centering
\includegraphics[width=0.92\textwidth]{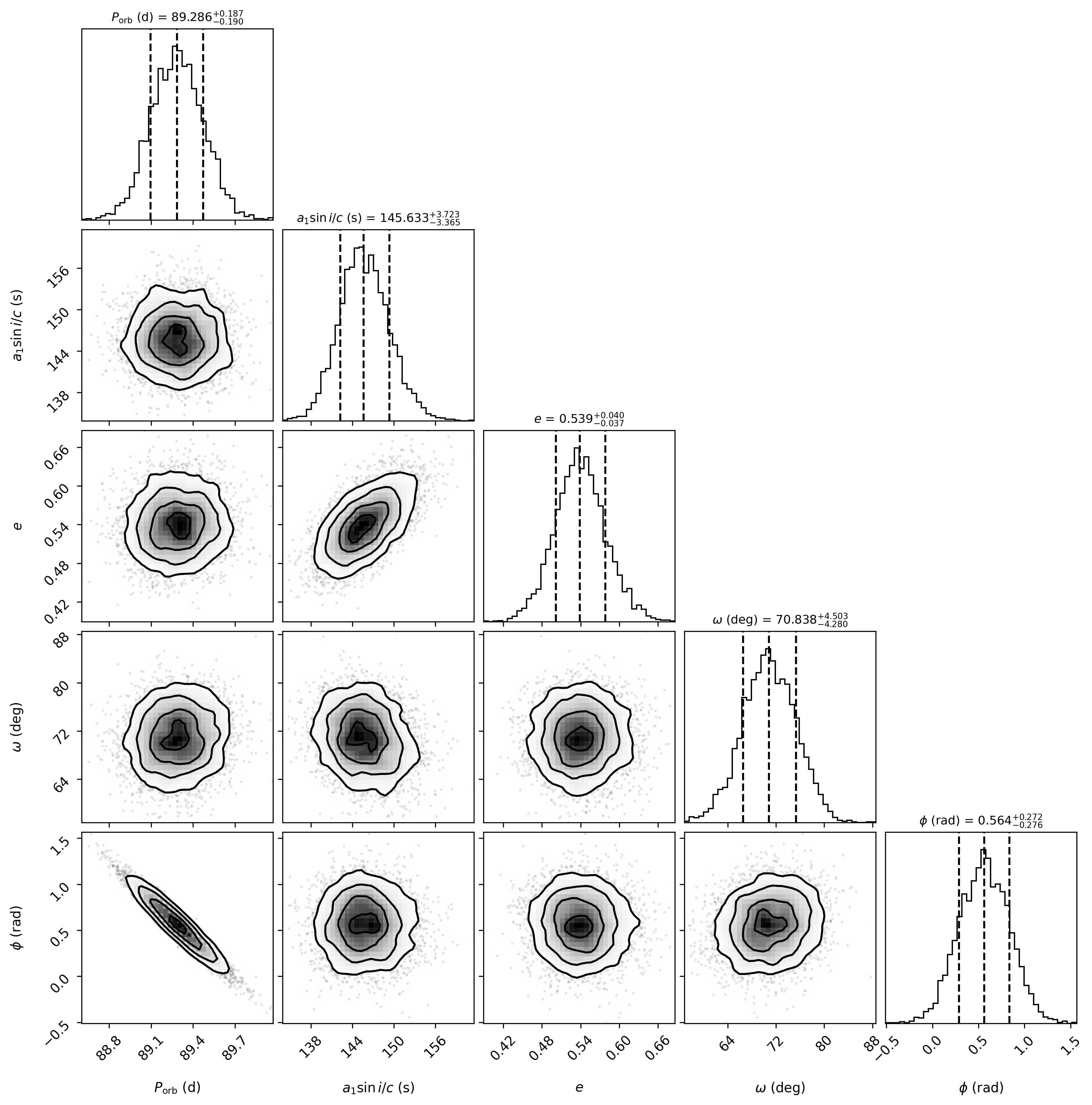}
\caption{Posterior distribution of the phase modulation orbital parameters. The diagonal panels show the marginalized distributions of $P_{\rm orb}$, $a_1\sin i/c$, $e$, $\omega$, and the orbital phase parameter $\phi$; the off diagonal panels show their pairwise covariances. Dashed lines mark the median and the 16th and 84th percentiles, and contours enclose progressively higher posterior density regions.}
\label{fig:orbit_corner}
\end{figure*}

The orbital parameter posterior is shown in Figure~\ref{fig:orbit_corner}. In addition to the parameters quoted above, the fit gives an argument of periastron \textbf{$\omega=70.8_{-4.3}^{+4.5}~{\rm deg}$} and an orbital phase parameter \textbf{$\phi=0.564_{-0.276}^{+0.272}~{\rm rad}$} at the adopted reference epoch. The posterior shows a strong covariance between $P_{\rm orb}$ and $\phi$, while $a_1\sin i/c$, $e$, and $\omega$ are comparatively well localized.

We additionally performed a FM analysis of the two strongest modes, $\nu_1=36.038013(9)~{\rm d^{-1}}$ and $\nu_2=34.29207(1)~{\rm d^{-1}}$, which show both first- and second-order orbital sidelobes, following \citet{ShibahashiKurtz2012} and the eccentric-orbit formalism of \citet{Shibahashi2015FM}. The frequencies in the two multiplets were constrained to $\nu_{i,m}=\nu_i+m\nu_{\rm orb}$ with a common orbital splitting. This gives $\nu_{\rm orb}=0.011191\pm0.000024~{\rm d^{-1}}$, corresponding to $P_{\rm FM}=89.36\pm0.19~{\rm d}$.

\begin{figure*}[t]
\centering
\includegraphics[width=0.95\textwidth]{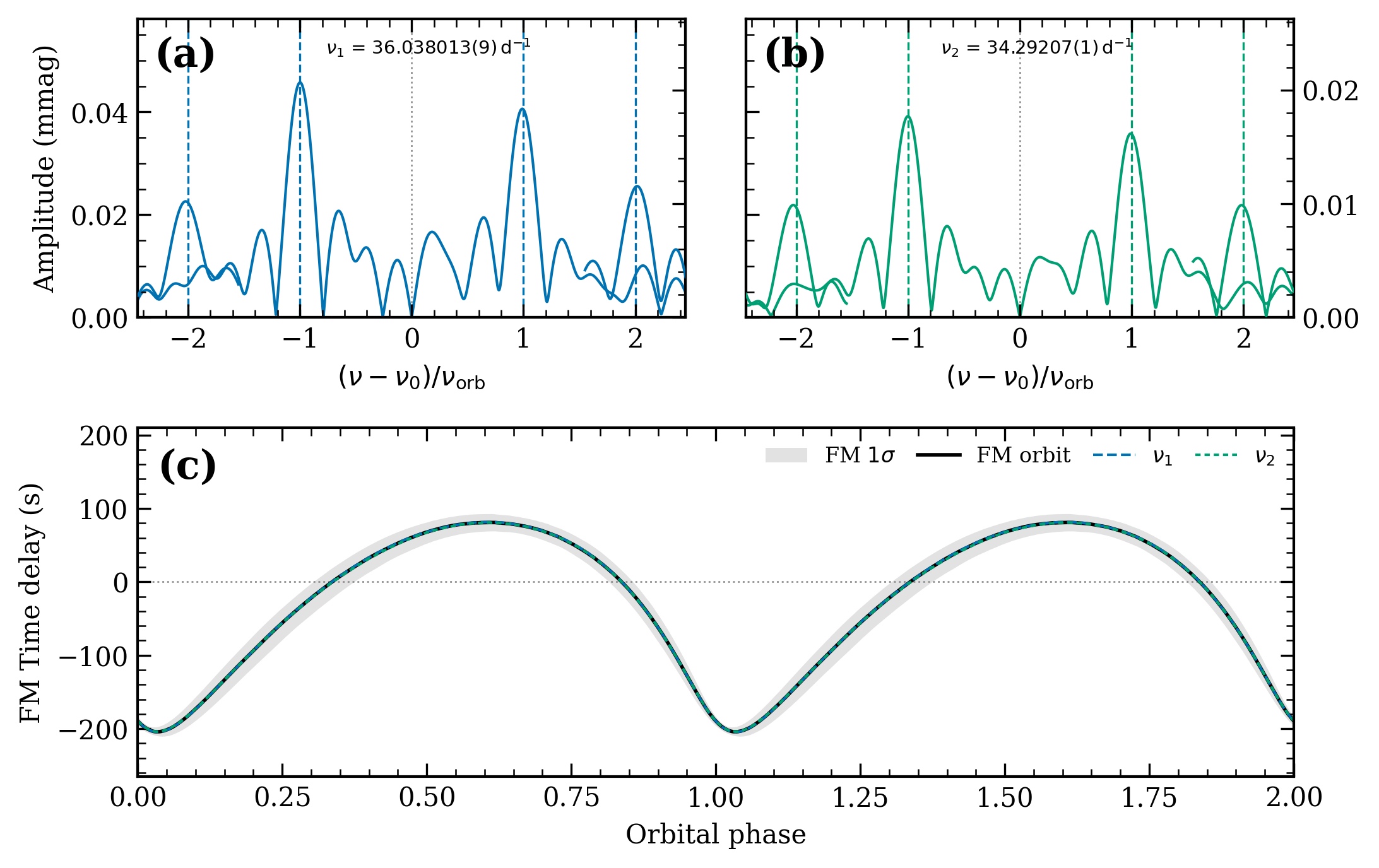}
\caption{FM analysis of TIC~160582982. (a) and (b) Fourier amplitude spectra around $\nu_1=36.038013(9)~{\rm d^{-1}}$ and $\nu_2=34.29207(1)~{\rm d^{-1}}$, respectively. The left ordinate shows the amplitudes after removal of the central pulsation, highlighting the first-order sidelobes. The right ordinate gives the smaller amplitude scale after removal of the central and first-order components, making the weaker second-order sidelobes visible. Dashed lines mark the expected orbital sidelobe positions. (c) Time-delay curve derived from the FM solution; the shaded region shows the 68\% credible interval.}
\label{fig:orbital_fm}
\end{figure*}

For the constrained FM fit, we adopted a common phase zero point defined by the $m=\pm1$ sidelobes of $\nu_1$. At this epoch, the fitted first-order sidelobes of $\nu_2$ are also consistent with the expected FM phase relation. Using the amplitudes and phases of the first- and second-order sidelobes, we obtain $a_1\sin i/c=148.4_{-4.3}^{+5.0}~{\rm s}$, $e_{\rm FM}=0.539_{-0.048}^{+0.052}$, and $\omega_{\rm FM}=60.1_{-5.9}^{+6.1}~{\rm deg}$. Figure~\ref{fig:orbital_fm} shows the two frequency multiplets and the corresponding FM time-delay curve. These results are consistent with the six-mode PM solution. We therefore adopt the PM orbit and use the FM result as an independent check.

No eclipses are detected in the TESS light curve. Because the system is wide and eccentric, this excludes only a narrow range of nearly edge-on inclinations, but it provides a useful geometric prior when assessing luminous main-sequence companion scenarios.

\section{SED Constraints and the Missing light Test}
\label{sec:sed}

To determine the properties of the visible star and to search for additional continuum light, we fitted the optical to mid infrared SED of TIC~160582982 with the \texttt{Python}-based tool \texttt{ASTROARIADNE}\footnote{\url{https://github.com/jvines/astroARIADNE}} \citep{Vines2022}. The fit used the archival photometry described in Section~\ref{sec:data}, Gaia parallax information as a distance constraint, and nested sampling with \texttt{dynesty} \citep{Speagle2020}. Several stellar atmosphere grids were considered, including PHOENIX, BT-Settl, Kurucz, and Castelli--Kurucz models \citep{Kurucz1993,CastelliKurucz2003,Husser2013,Allard2012}. Bayesian model averaging was enabled; the largest posterior weight is assigned to Castelli--Kurucz type models, consistent with an A-type photosphere.

\begin{figure}[t]
\centering
\includegraphics[width=0.48\textwidth]{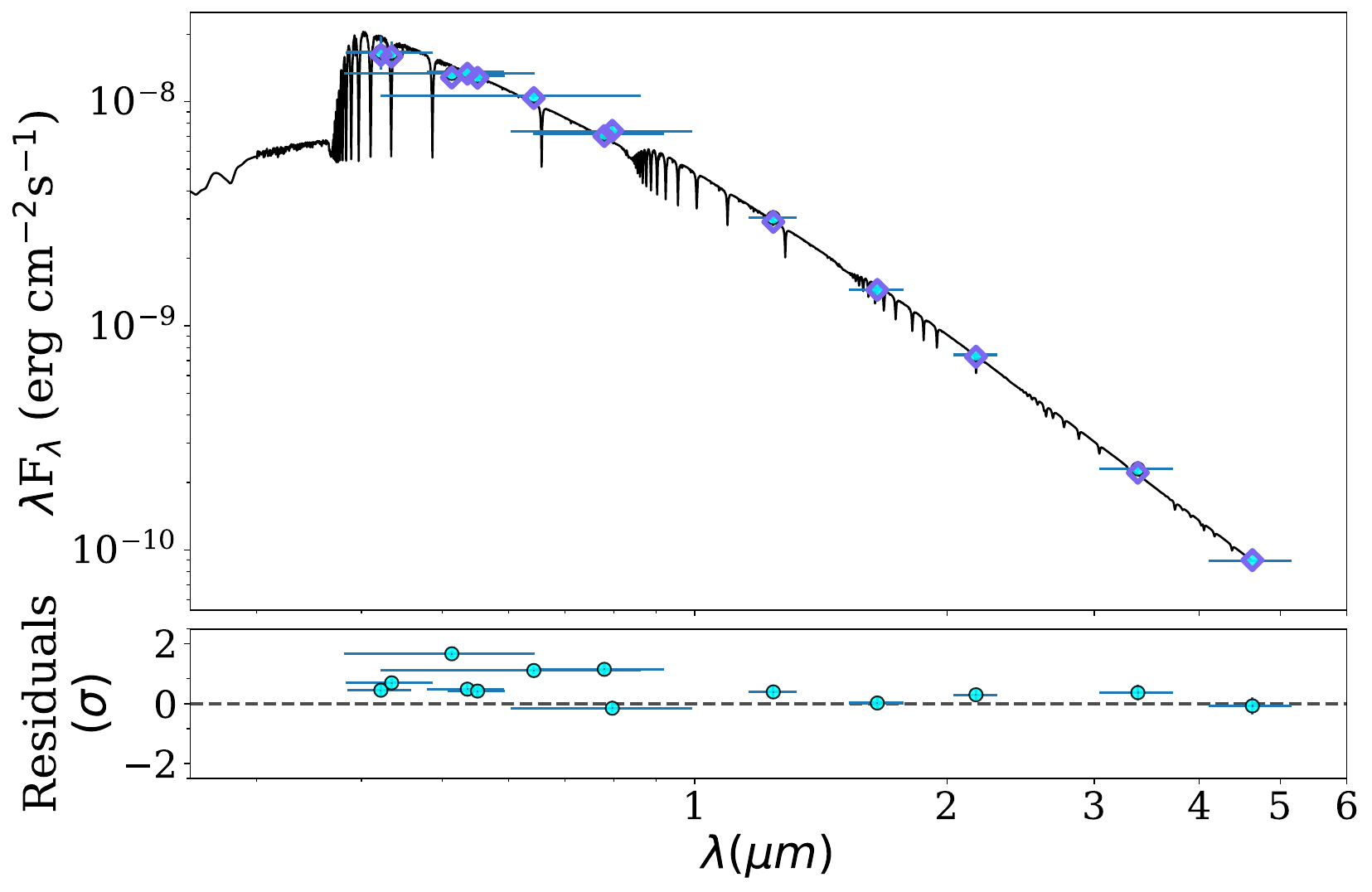}
\caption{Optical to mid infrared SED of TIC~160582982. The observed fluxes are fitted with a single stellar atmosphere, and the lower panel shows the residuals in units of the observational uncertainty. The absence of a WISE infrared excess indicates that the system is not dominated by warm dust emission or circumstellar reprocessing.}
\label{fig:sed_fitting}
\end{figure}

Figure~\ref{fig:sed_fitting} shows that the observed fluxes are well described by a single stellar atmosphere. The best fitting parameters are $T_{\rm eff}=8560\pm60~{\rm K}$, $\log g=3.78\pm0.10$, $[{\rm Fe/H}]=-0.08\pm0.09$, $D=262.7\pm1.8~{\rm pc}$, $R_1=2.79\pm0.03~R_\odot$, and $A_V=0.037\pm0.011~{\rm mag}$. The corresponding luminosity is $L=37.6\pm1.3~L_\odot$, and the SED-based spectral type is A3V. These formal uncertainties reflect the adopted atmosphere grids, photometric errors, and model assumptions, and should not be interpreted as the full systematic uncertainty for a rapidly rotating A star. We therefore use the SED radius as an external radius constraint in the rotating isochrone inference, rather than as an exact radius in the dynamical analysis.

No significant infrared excess is detected in the WISE bands. The W1 and W2 fluxes follow the stellar photosphere within the photometric uncertainties, arguing against warm dust emission or strong circumstellar reprocessing. More importantly, the broadband SED does not require an additional luminous stellar component.

We performed a first order, hypothesis conditional missing light test to estimate the continuum flux expected if the companion were an ordinary main-sequence star. The primary was fixed to the single star SED solution, $T_{\rm eff,1}=8556~{\rm K}$ and $R_1=2.789~R_\odot$. For a grid of representative main-sequence companions spanning $1.70$--$3.00~M_\odot$, including the dynamical lower limit region, we assigned approximate $T_{\rm eff,2}$ and $R_2$ values guided by empirical main-sequence relations \citep{PecautMamajek2013,Eker2018}, and estimated the monochromatic continuum contribution at the effective wavelength of each observed band using $F_\nu\propto R^2B_\nu(T)$. This calculation is a forward diagnostic under the single primary SED solution, not a posterior constraint on $M_2$ and not a substitute for a simultaneous two star SED fit. In particular, it is evaluated independently of the non eclipse conditioned $M_2$--$i$ posterior.

The result is shown in Appendix~\ref{app:missing_light}. Even near the conservative lower mass end, a $1.70$--$1.86~M_\odot$ main-sequence companion would contribute $\simeq18$--$26\%$ of the Gaia $G$-band flux and $\simeq21$--$27\%$ of the WISE W1 flux. At $2.30~M_\odot$, included only as a representative higher mass case, the estimated contribution is $\simeq43\%$ in Gaia $G$ and $\simeq40\%$ in both 2MASS $K_s$ and WISE W1. Such continuum light is difficult to reconcile with the observed optical to mid infrared SED, which is already well described by a single A-type photosphere. This test effectively rules out an ordinary unobscured main-sequence companion.

\section{Primary star constraints and Companion Mass}
\label{sec:companion_mass}

\begin{figure}[t]
\centering
\includegraphics[width=0.48\textwidth]{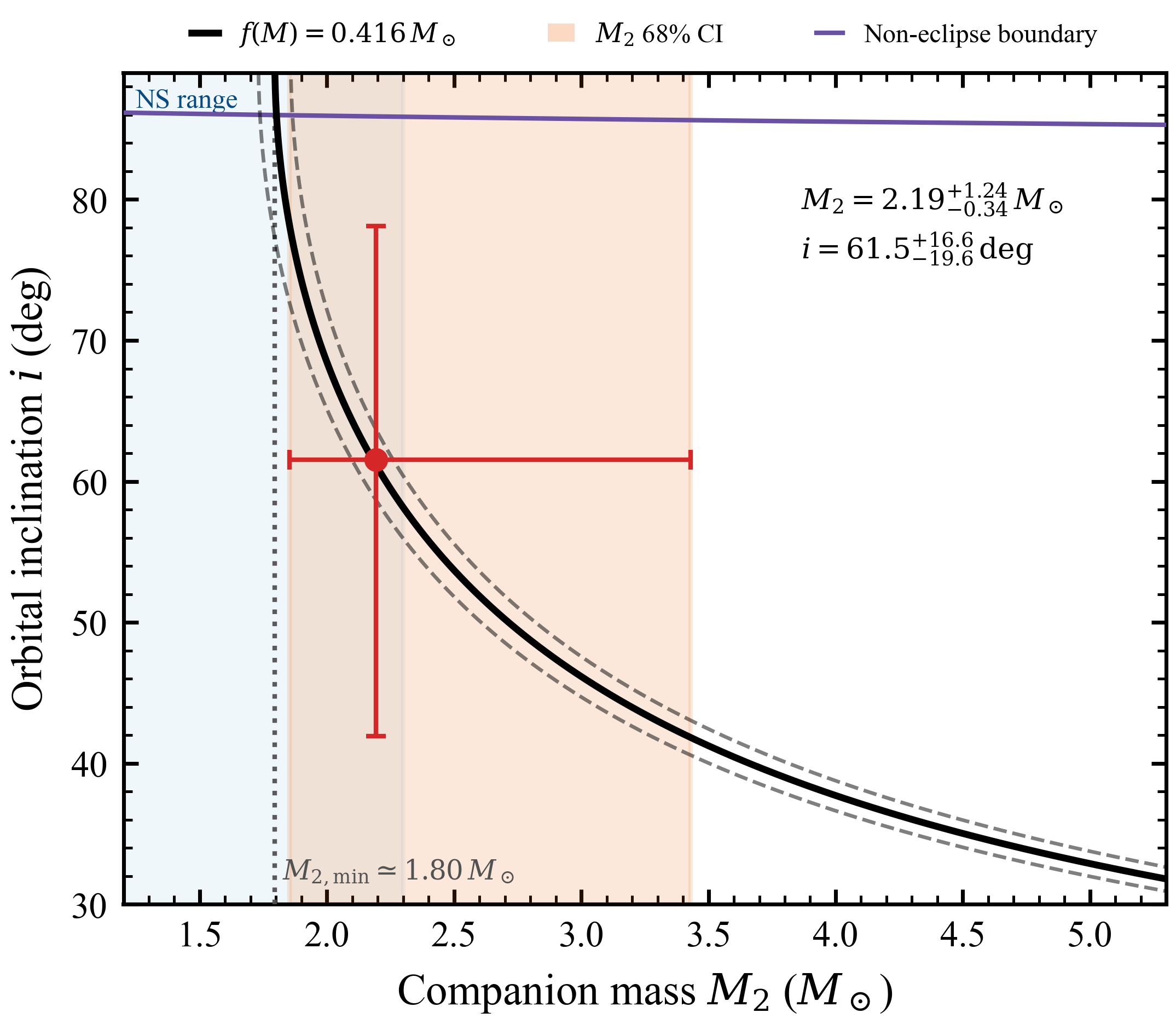}
\caption{Companion mass constraint for TIC~160582982. The black curve shows the $M_2$--$i$ relation required by the phase modulation mass function, $f(M)=0.416~M_\odot$, with dashed curves indicating the $\pm1\sigma$ range. The orange band marks the 68\% credible interval obtained only under the luminous main sequence hypothesis and its non-eclipse prior, $M_2=2.19_{-0.34}^{+1.24}~M_\odot$. The blue region indicates a neutron star reference mass range, while the vertical line marks the formal edge-on minimum companion mass allowed by the mass function, $M_{2,\min}\simeq1.80~M_\odot$. The purple curve shows the non-eclipsing boundary for the adopted luminous main-sequence companion radius estimate.}
\label{fig:m2_inclination}
\end{figure}

The SED fit and Gaia DR3 astrophysical parameters constrain the properties of the visible primary, while the ExoFOP-TESS classification diagnostics indicate substantial rotational broadening. Therefore, we used the MIST v1.2 rotating isochrones computed with an initial $v/v_{\rm crit}=0.4$ to construct the primary star prior \citep{Dotter2016,Choi2016}. The grid was likelihood weighted using the observed $T_{\rm eff}$, $\log g$, and SED-based radius. We did not impose a strong metallicity constraint because metallicity estimates for a rapidly rotating A star can be affected by systematic errors.

The rotating MIST inference gives $M_1=1.93\pm0.19~M_\odot$, $R_1=2.65\pm0.23~R_\odot$, and an age of $\tau=0.89_{-0.26}^{+0.23}~{\rm Gyr}$. These values are consistent, at the precision required here, with the SED radius and with the Gaia DR3 astrophysical parameters \citep{GaiaCollaboration2023,Creevey2023}. Gaia DR3 gives $T_{\rm eff}=8538~{\rm K}$, $\log g=3.778$, $M_{\rm Gaia}=2.23~M_\odot$, and $R_{\rm Gaia}=2.69~R_\odot$, with a FLAME radius of $R_{\rm FLAME}=2.76~R_\odot$. Given these limitations, we adopted the conservative rotating MIST values, $M_1=1.93\pm0.19~M_\odot$ and $R_1=2.65\pm0.23~R_\odot$, as Gaussian priors in the companion mass inference.

We first obtained the hypothesis independent dynamical lower limit by combining \textbf{$f(M)=0.416_{-0.028}^{+0.032}~M_\odot$} with the primary mass prior at $i=90^\circ$, giving $M_{2,\min}\simeq1.80~M_\odot$. We then performed a separate, explicitly conditional inference for the luminous main sequence hypothesis. The Markov chain Monte Carlo (MCMC) model sampled $\log M_2$, $\cos i$, $M_1$, and $R_1$, using an isotropic orientation prior and Gaussian constraints on $f(M)$, $M_1$, and $R_1$ \citep{ForemanMackey2013}. The non-eclipsing condition was imposed geometrically at periastron for a luminous companion. For this purpose, we adopted an approximate main-sequence radius relation, $R_2\simeq M_2^{0.75}~R_\odot$, broadly consistent with empirical main-sequence mass--radius relations in this mass range \citep{Eker2018}. This radius relation is used only within the luminous main sequence hypothesis; for a compact companion the non-eclipse constraint would be negligible.

Figure~\ref{fig:m2_inclination} shows the resulting constraint in the $M_2$--$i$ plane. For the adopted primary mass, the mass function alone gives a formal edge-on lower limit of $M_{2,\min}\simeq1.80~M_\odot$. This limiting value is useful as a dynamical reference, although an exactly edge-on configuration would be disfavored for a luminous main-sequence companion by the absence of eclipses. Within the luminous main sequence hypothesis, including the primary star priors and the non-eclipsing prior yields $M_2=2.19_{-0.34}^{+1.24}~M_\odot$ and $i=61.5_{-19.6}^{+16.6}~{\rm deg}$. The posterior is asymmetric because lower inclinations allow substantially
larger companion masses, while the non-eclipsing condition mainly removes the highest inclination tail for luminous companion configurations. This posterior is conditional on the hypothesis that the companion is a single main-sequence star.

\section{Archival Spectroscopic Context}
\label{sec:spectroscopy}

We inspected the archival TRES information available from the ExoFOP-TESS database\footnote{\url{https://exofop.ipac.caltech.edu/tess/target.php?id=160582982}} as qualitative spectroscopic context. The classification products are
consistent with a rapidly rotating A-type primary and show no obvious second cross-correlation peak or clearly separated secondary absorption component. The archived entries do not provide an orbital radial velocity
series and therefore cannot independently validate the phase modulation orbit or further constrain the mass function.

The broad lines of the primary limit the diagnostic value of the available classification products. A weaker secondary component could remain undetected depending on its flux contribution, velocity separation, and
rotational broadening. We therefore conclude only that no conspicuous secondary spectral component is present; quantitative limits on companion light require phase-resolved high resolution spectroscopy.

\section{The Nature of the Dynamical Companion}
\label{sec:companion_nature}

The phase modulation orbit constrains the mass associated with the outer orbit but does not establish whether that mass belongs to a single object. For the adopted primary star prior, the mass function gives a formal edge-on lower limit of $M_{2,\min}\simeq1.80~M_\odot$. If the companion is
a single main-sequence star near this limit, the first order continuum calculation in Appendix~\ref{app:missing_light} predicts a contribution of approximately 20--25\% in the optical and near-infrared bands. The absence of an obvious additional component in the broadband SED and archival TRES diagnostics therefore strongly disfavors this single-main-sequence interpretation.

A hierarchical triple provides a viable alternative. In this configuration, the 89.29-d orbit would describe the motion of the $\delta$ Scuti star around the barycenter of an unresolved inner binary. At the formal dynamical lower limit, the required mass could be supplied
by two stars of approximately $0.9~M_\odot$ each. Because the main-sequence mass--luminosity relation is nonlinear, their combined luminosity is substantially lower than that of a single $1.8~M_\odot$ star. Extending the first order continuum calculation to
such an equal mass inner binary gives companion light fractions of approximately 3.8\% in Gaia $G$, 8.3\% in 2MASS $K_s$, and 9.1\% in WISE W1. The predicted contribution increases as the inner binary mass
ratio departs from unity, but a near equal mass pair could remain much less conspicuous in the current SED than a single star with the same total mass. The inner orbit would need to be sufficiently compact for the system to remain dynamically hierarchical, but its period and orientation cannot be constrained with the present data.

A compact remnant would likewise contribute little or no optical and infrared continuum. The present photometric and spectroscopic constraints therefore do not distinguish between an unresolved inner binary and a compact remnant, while an ordinary unobscured single main-sequence companion is strongly disfavored. The archival TRES classification products show no conspicuous secondary component, but the rotationally broadened primary spectrum does not provide a quantitative limit on weak companion light.

Multi-epoch, high resolution spectroscopy covering the 89.29-d orbit offers the most direct way to distinguish these possibilities. Such observations would independently test the phase modulation orbit and
search for weak spectral features moving with either a single companion or the components of an inner binary. Future Gaia astrometry may provide a complementary constraint on the orbital inclination and hence on the
total companion mass.

\section{A Low-frequency Signal with Orbital-phase-dependent Variability}
\label{sec:lowfreq}

In addition to the high frequency pressure modes used for the phase modulation orbit, TIC~160582982 shows a low frequency signal at \textbf{$\nu_{\rm low}=1.8383(1)~{\rm d^{-1}}$}, corresponding to $P_{\rm low}=0.54398~{\rm d}$. This frequency lies in the usual $\gamma$ Doradus/g-mode or rotation modified gravito-inertial regime of A/F-type pulsators \citep{Aerts2010,Ballot2010}. For $P_{\rm orb}=89.29~{\rm d}$, the orbital frequency is \textbf{$\nu_{\rm orb}=0.011199~{\rm d^{-1}}$}, and \textbf{$\nu_{\rm low}/\nu_{\rm orb}=164.14$}. This proximity to a high order orbital harmonic is suggestive, as in heartbeat systems where tidally excited oscillations often appear at integer multiples of the orbital frequency \citep[e.g.,][]{Welsh2011,FullerLai2012,Fuller2017}, but it is not by itself sufficient to identify a tidally excited oscillation \citep{Kumar1995}.

\begin{figure*}[t]
\centering
\includegraphics[width=0.95\textwidth]{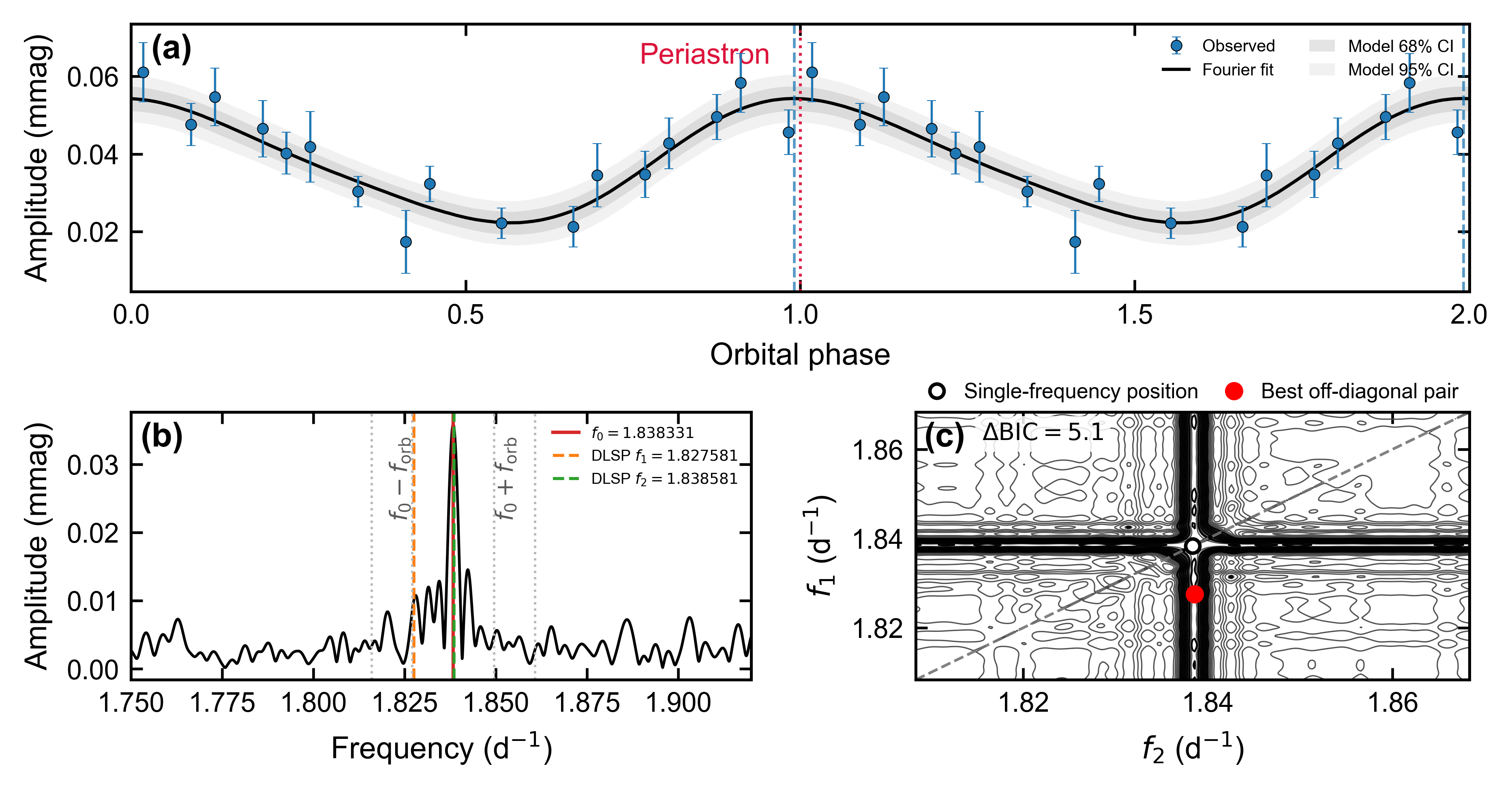}
\caption{Diagnostics of the low frequency signal at $\nu_{\rm low}=1.8383(1)~{\rm d^{-1}}$}.
(a) Orbital phase dependent amplitude variation from fixed frequency sliding window fits. Points show phase binned amplitudes and are not connected. The black curve shows a second order Fourier fit at $P_{\rm orb}=89.29~{\rm d}$. Red dotted vertical lines mark periastron passages, and blue dashed vertical lines mark the fitted amplitude maximum.
(b) Amplitude spectrum around $\nu_{\rm low}$.
(c) Double-frequency search around $\nu_{\rm low}$. The best off diagonal solution has $\Delta \nu=0.01100~{\rm d^{-1}}$, close to $\nu_{\rm orb}=0.01120~{\rm d^{-1}}$.

\label{fig:lowfreq_diagnostics}
\end{figure*}

We measured the amplitude of the $1.8383(1)~{\rm d^{-1}}$ signal by fitting sine and cosine terms at this fixed frequency, together with a linear local trend, in 20-d windows advanced by 10~d. The resulting 33 amplitudes were folded on the orbital period, with phase zero at periastron, and grouped into 28 equal phase bins, 17 of which contain measurements. Figure~\ref{fig:lowfreq_diagnostics} shows that the amplitude varies with orbital phase. A second order Fourier fit gives $A_0=0.0378~{\rm mmag}$ and $A_1=0.0155~{\rm mmag}$, corresponding to $A_1/A_0\simeq0.41$. The fitted amplitude maximum occurs at $\Phi_{\rm max}=0.991$, approximately $0.8~{\rm d}$ before periastron, and is consistent with periastron within the formal uncertainty. Because the windows overlap, this uncertainty should be interpreted with caution.

We tested whether the apparent orbital modulation could instead be reproduced by a second nearby sinusoidal component. A double Lomb--Scargle periodogram (DLSP) \citep{Scargle2026} gives a best frequency separation of $\Delta \nu=0.01100~{\rm d^{-1}}$, close to $\nu_{\rm orb}=0.01120~{\rm d^{-1}}$. However, the improvement over the single frequency model is modest, with $\Delta{\rm BIC}=5.1$ \citep{Schwarz1978}. Because orbital amplitude modulation can also produce sidebands separated by $\nu_{\rm orb}$, this result does not provide strong evidence for an independent nearby mode. Unrelated two-mode beating is therefore unlikely to be the main cause of the observed orbital phase dependence.

A rotational interpretation remains possible because the archival TRES classifications indicate strong line broadening. However, the available classification products do not provide a sufficiently robust $v\sin i$ measurement for a quantitative rotation-frequency or spin-inclination inference. The proximity to a high-order orbital harmonic, the coherent orbital-phase dependence, and the absence of strong DLSP evidence for an independent close mode suggest a possible tidal connection rather than unrelated beating. Photometry alone cannot establish whether the sideband is a distinct mode or the Fourier signature of modulation, so we retain the physically conservative classification of the $1.8383(1)~{\rm d^{-1}}$ signal as a tidally affected g-mode or gravito-inertial-mode candidate rather than a confirmed tidally excited oscillation.

\section{Summary and Discussion}
\label{sec:summary}

We have identified TIC~160582982 as a long-period, eccentric
$\delta$ Scuti pulsating binary. Six pressure modes yield a
phase modulation orbit with
$P_{\rm orb}=89.29\pm0.19~{\rm d}$,
$e=0.539_{-0.037}^{+0.040}$,
$a_1\sin i/c=145.6_{-3.4}^{+3.7}~{\rm s}$, and $f(M)=0.416_{-0.028}^{+0.032}~M_\odot$. The independent FM analysis of the $\nu_{1}$ and $\nu_{2}$ quintuplets gives $P_{\rm FM}=89.36\pm0.19~{\rm d}$, $a_1\sin i/c=148.4_{-4.3}^{+5.0}~{\rm s}$, and $e_{\rm FM}=0.539_{-0.048}^{+0.052}$, in agreement with the adopted PM orbit. For the adopted rotating MIST primary mass of
$M_1=1.93\pm0.19~M_\odot$, the mass function gives a formal edge-on
minimum companion mass of
$M_{2,\min}\simeq1.80~M_\odot$. Under the additional assumption that the
companion is a single main-sequence star, the primary star priors and non-eclipse condition yield $M_2=2.19_{-0.34}^{+1.24}~M_\odot$ and $i=61.5_{-19.6}^{+16.6}~{\rm deg}$.

The optical-to-mid infrared SED is well described by a single A-type
atmosphere, and the archival TRES classification products show no
conspicuous secondary component. A first order continuum calculation
indicates that a single main-sequence companion near the dynamical lower
limit could contribute approximately 20--25\% of the optical and
near-infrared light. This strongly disfavors the single-main-sequence interpretation.

The dynamical mass need not belong to a single object. At the formal lower
limit, an unresolved inner binary containing two
$0.9~M_\odot$ main-sequence stars would contribute only about 4\% of the
Gaia $G$-band light and 8--9\% in 2MASS $K_s$ and WISE W1 under the same
first order continuum approximation. A near equal mass hierarchical
configuration could therefore satisfy the dynamical constraint while
remaining substantially less prominent than a single
$1.8~M_\odot$ star. A compact remnant would likewise contribute little or no optical continuum. The present observations therefore strongly disfavor an ordinary unobscured single main-sequence companion, while an unresolved inner binary and a compact object such as a neutron star or a stellar-mass black hole remain viable configurations.

Flux contamination is unlikely to account for the low-frequency variability. Across the 12 sectors, the SPOC CROWDSAP values range from 0.9958 to 0.9976, corresponding to estimated contaminating flux fractions of only 0.24--0.42\%. TASOC independently reports a contamination fraction of 0.17\%. These consistently low values make a nearby contaminating source an unlikely origin of $1.8383(1)~{\rm d^{-1}}$. This signal shows orbital-phase-dependent amplitude variability, while the preferred two-frequency separation lies close to $\nu_{\rm orb}$ but provides only a modest improvement over a single-frequency model. We therefore retain the conservative interpretation of the $1.8383(1)~{\rm d^{-1}}$ signal as a tidally affected g-mode or gravito-inertial-mode candidate rather than a confirmed tidally excited oscillation.

TIC~160582982 is an interesting target for follow-up. The phase modulation orbit
corresponds to an angular projected primary orbit of
$\alpha_1\sin i\simeq1.1~{\rm mas}$ at the Gaia distance. Multi-epoch
high resolution spectroscopy covering the 89.29-d orbit can independently
test the timing solution and search for weak moving spectral features from
a single companion or an unresolved inner pair. Future Gaia astrometry may
provide a complementary constraint on the orbital inclination and total
companion mass. These observations are needed to determine the
architecture and nature of the dynamical companion.

\begin{acknowledgments}
We are grateful to the anonymous reviewer for the professional and thorough suggestions. Chenglong Lv acknowledges support from the National Natural Science Foundation of China (NSFC; No. 1250030315) and the Natural Science Foundation of Xinjiang Uygur Autonomous Region (No. 2025D01B176). Ali Esamdin is supported by the Tian-shan Talent Training Program (No. 2023TSYCLJ0053). This work is also supported by the Xinjiang Uygur Autonomous Region Introduced Project ``Tianchi Talent.'' JPG acknowledges financial support from project PID2023-149439NB-C42 from the `Proyectos de Generaci\'on de Conocimiento' and from the Severo Ochoa grant CEX2021-001131-S funded by MICIU/AEI/10.13039/501100011033 and FEDER, EU.

This paper includes data collected with the \textit{TESS} mission, obtained from the Mikulski Archive for Space Telescopes (MAST) at the Space Telescope Science Institute (STScI). Funding for the \textit{TESS} mission is provided by the NASA Explorer Program. STScI is operated by the Association of Universities for Research in Astronomy, Inc., under NASA contract NAS~5-26555. The specific TESS SPOC 2-min observations analyzed in this work can be accessed from MAST via \dataset[doi:10.17909/wceq-v793]{https://doi.org/10.17909/wceq-v793}. This work also presents results from the European Space Agency (ESA) space mission Gaia. Gaia data are being processed by the Gaia Data Processing and Analysis Consortium (DPAC). Funding for the DPAC is provided by national institutions, in particular the institutions participating in the Gaia MultiLateral Agreement (MLA). The Gaia mission website is \url{https://www.cosmos.esa.int/gaia}. The Gaia archive website is \url{https://archives.esac.esa.int/gaia}.

\software{
\texttt{astropy} \citep{Astropy2013,Astropy2018,Astropy2022},
\texttt{matplotlib} \citep{Hunter2007},
\texttt{numpy} \citep{Harris2020},
\texttt{scipy} \citep{Virtanen2020},
\texttt{emcee} \citep{ForemanMackey2013},
\texttt{Lightkurve} \citep{Lightkurve2018},
\texttt{FELIX} \citep{Charpinet2010},
\texttt{Maelstrom} \citep{Hey2020},
\texttt{ASTROARIADNE} \citep{Vines2022}
}

\end{acknowledgments}

\appendix

\twocolumngrid

\section{Frequency List}
\label{app:frequency_list}

Table~\ref{tab:frequencies} lists the 128 frequencies extracted from the combined SPOC light curve described in Section~\ref{sec:data}.

\begin{deluxetable*}{cccccccccc}
\tabletypesize{\scriptsize}
\tablewidth{\textwidth}

\tablecaption{Extracted Pulsation Frequencies and Fitted Sinusoidal Parameters for TIC~160582982\label{tab:frequencies}}

\tablehead{
\colhead{ID} &
\colhead{Fre (${\rm d^{-1}}$)} &
\colhead{Amp (mmag)} &
\colhead{Phase} &
\colhead{S/N} &
\colhead{ID} &
\colhead{Fre (${\rm d^{-1}}$)} &
\colhead{Amp (mmag)} &
\colhead{Phase} &
\colhead{S/N}
}
\startdata
$\nu_{1}$  & 36.038013(9) & 0.296(2) & 0.299(2) & 171.9 & $\nu_{65}$  & 0.2123(3)  & 0.016(3) & 0.52(6) & 5.6 \\
$\nu_{2}$  & 34.29207(1)  & 0.262(2) & 0.603(2) & 154.7 & $\nu_{66}$  & 18.5032(2) & 0.016(2) & 0.75(4) & 8.9 \\
$\nu_{3}$  & 24.14168(1)  & 0.256(2) & 0.808(2) & 150.3 & $\nu_{67}$  & 41.8848(2) & 0.016(2) & 0.81(4) & 8.8 \\
$\nu_{4}$  & 39.87724(2)  & 0.125(2) & 0.433(4) & 70.1  & $\nu_{68}$  & 49.4235(2) & 0.016(2) & 0.49(4) & 8.7 \\
$\nu_{5}$  & 43.19855(3)  & 0.113(2) & 0.138(5) & 59.2  & $\nu_{69}$  & 35.3861(2) & 0.015(2) & 0.86(4) & 8.5 \\
$\nu_{6}$  & 32.10800(2)  & 0.111(2) & 0.242(5) & 64.6  & $\nu_{70}$  & 36.0604(2) & 0.015(2) & 0.39(4) & 8.6 \\
$\nu_{7}$  & 39.79930(4)  & 0.074(2) & 0.973(8) & 41.8  & $\nu_{71}$  & 42.0034(2) & 0.015(2) & 0.51(4) & 8.4 \\
$\nu_{8}$  & 21.57594(4)  & 0.072(2) & 0.734(7) & 42.3  & $\nu_{72}$  & 44.0944(2) & 0.015(2) & 0.62(4) & 8.5 \\
$\nu_{9}$  & 29.85261(4)  & 0.070(2) & 0.823(8) & 39.3  & $\nu_{73}$  & 44.7793(2) & 0.015(2) & 0.02(4) & 8.4 \\
$\nu_{10}$ & 31.83890(4)  & 0.066(2) & 0.743(8) & 38.7  & $\nu_{74}$  & 58.3667(2) & 0.015(2) & 0.24(4) & 9.1 \\
$\nu_{11}$ & 36.79738(4)  & 0.062(2) & 0.251(9) & 36.1  & $\nu_{75}$  & 36.0361(2) & 0.014(2) & 0.20(4) & 8.0 \\
$\nu_{12}$ & 33.69973(5)  & 0.059(2) & 0.047(9) & 34.5  & $\nu_{76}$  & 36.6399(2) & 0.014(2) & 0.22(4) & 8.1 \\
$\nu_{13}$ & 36.02687(5)  & 0.054(2) & 0.59(1)  & 31.1  & $\nu_{77}$  & 39.7711(2) & 0.014(2) & 0.38(4) & 7.9 \\
$\nu_{14}$ & 36.04917(6)  & 0.049(2) & 0.49(1)  & 28.5  & $\nu_{78}$  & 44.0887(2) & 0.014(2) & 0.68(4) & 7.9 \\
$\nu_{15}$ & 26.58873(6)  & 0.047(2) & 0.35(1)  & 27.8  & $\nu_{79}$  & 53.0928(2) & 0.014(2) & 0.26(4) & 7.9 \\
$\nu_{16}$ & 34.28083(6)  & 0.045(2) & 0.07(1)  & 26.8  & $\nu_{80}$  & 18.3906(2) & 0.013(2) & 0.64(4) & 7.2 \\
$\nu_{17}$ & 37.14240(6)  & 0.044(2) & 0.23(1)  & 25.7  & $\nu_{81}$  & 27.2461(2) & 0.013(2) & 0.82(4) & 7.3 \\
$\nu_{18}$ & 40.04692(7)  & 0.043(2) & 0.59(1)  & 23.4  & $\nu_{82}$  & 28.6701(2) & 0.013(2) & 0.06(4) & 8.0 \\
$\nu_{19}$ & 34.30331(6)  & 0.041(2) & 0.67(1)  & 24.2  & $\nu_{83}$  & 32.7594(2) & 0.013(2) & 0.86(4) & 7.2 \\
$\nu_{20}$ & 42.90420(7)  & 0.040(2) & 0.12(1)  & 21.5  & $\nu_{84}$  & 39.7886(2) & 0.013(2) & 0.48(4) & 7.4 \\
$\nu_{21}$ & 46.25449(7)  & 0.040(2) & 0.41(1)  & 21.7  & $\nu_{85}$  & 41.3197(2) & 0.013(2) & 0.12(4) & 7.1 \\
$\nu_{22}$ & 45.85701(8)  & 0.038(2) & 0.90(1)  & 20.6  & $\nu_{86}$  & 43.6666(2) & 0.013(2) & 0.48(4) & 7.2 \\
$\nu_{23}$ & 20.29593(7)  & 0.037(2) & 0.57(1)  & 23.1  & $\nu_{87}$  & 45.8366(2) & 0.013(2) & 0.42(5) & 7.0 \\
$\nu_{24}$ & 33.92978(7)  & 0.037(2) & 0.28(1)  & 21.6  & $\nu_{88}$  & 51.6348(2) & 0.013(2) & 0.01(4) & 7.4 \\
$\nu_{25}$ & 1.8383(1)    & 0.036(2) & 0.95(2)  & 16.2  & $\nu_{89}$  & 29.6485(2) & 0.012(2) & 0.04(5) & 6.6 \\
$\nu_{26}$ & 27.07590(8)  & 0.036(2) & 0.18(1)  & 20.5  & $\nu_{90}$  & 33.6885(2) & 0.012(2) & 0.45(4) & 7.1 \\
$\nu_{27}$ & 45.84729(8)  & 0.036(2) & 0.03(2)  & 19.7  & $\nu_{91}$  & 35.2227(2) & 0.012(2) & 0.78(4) & 7.3 \\
$\nu_{28}$ & 18.63113(9)  & 0.032(2) & 0.17(2)  & 18.3  & $\nu_{92}$  & 36.0157(2) & 0.012(2) & 0.74(5) & 6.9 \\
$\nu_{29}$ & 22.89117(9)  & 0.032(2) & 0.09(2)  & 18.0  & $\nu_{93}$  & 38.3852(2) & 0.012(2) & 0.87(4) & 7.2 \\
$\nu_{30}$ & 36.68762(9)  & 0.032(2) & 0.10(2)  & 18.3  & $\nu_{94}$  & 38.4154(2) & 0.012(2) & 0.45(4) & 7.1 \\
$\nu_{31}$ & 24.13050(9)  & 0.031(2) & 0.16(2)  & 18.2  & $\nu_{95}$  & 39.8719(2) & 0.012(2) & 0.21(5) & 6.8 \\
$\nu_{32}$ & 45.82542(9)  & 0.031(2) & 0.40(2)  & 16.6  & $\nu_{96}$  & 45.8449(2) & 0.012(2) & 0.09(5) & 6.8 \\
$\nu_{33}$ & 2.4174(1)    & 0.030(2) & 0.47(2)  & 14.0  & $\nu_{97}$  & 55.6690(2) & 0.012(2) & 0.60(5) & 6.6 \\
$\nu_{34}$ & 39.87960(9)  & 0.030(2) & 0.80(2)  & 16.8  & $\nu_{98}$  & 57.2020(2) & 0.012(2) & 0.40(4) & 7.5 \\
$\nu_{35}$ & 36.5166(1)   & 0.029(2) & 0.45(2)  & 16.4  & $\nu_{99}$  & 21.1935(2) & 0.011(2) & 0.65(5) & 6.7 \\
$\nu_{36}$ & 24.1529(1)   & 0.028(2) & 0.91(2)  & 16.5  & $\nu_{100}$ & 21.5647(2) & 0.011(2) & 0.25(5) & 6.5 \\
$\nu_{37}$ & 29.8367(1)   & 0.028(2) & 0.35(2)  & 15.9  & $\nu_{101}$ & 29.8636(2) & 0.011(2) & 0.41(5) & 6.5 \\
$\nu_{38}$ & 43.8755(1)   & 0.028(2) & 0.29(2)  & 15.5  & $\nu_{102}$ & 30.1427(3) & 0.011(2) & 0.04(5) & 6.1 \\
$\nu_{39}$ & 39.9911(1)   & 0.026(2) & 0.92(2)  & 14.6  & $\nu_{103}$ & 33.8666(2) & 0.011(2) & 0.42(5) & 6.4 \\
$\nu_{40}$ & 37.8023(1)   & 0.024(2) & 0.21(2)  & 13.8  & $\nu_{104}$ & 34.8644(2) & 0.011(2) & 0.98(5) & 6.7 \\
$\nu_{41}$ & 43.1872(1)   & 0.023(2) & 0.82(3)  & 12.0  & $\nu_{105}$ & 41.3262(3) & 0.011(2) & 0.32(5) & 5.9 \\
$\nu_{42}$ & 47.5206(1)   & 0.023(2) & 0.45(2)  & 13.1  & $\nu_{106}$ & 42.3996(3) & 0.011(2) & 0.66(6) & 5.8 \\
$\nu_{43}$ & 46.2520(1)   & 0.022(2) & 0.70(3)  & 11.7  & $\nu_{107}$ & 42.9713(3) & 0.011(2) & 0.99(6) & 5.8 \\
$\nu_{44}$ & 21.8663(1)   & 0.021(2) & 0.17(3)  & 12.7  & $\nu_{108}$ & 49.0513(3) & 0.011(2) & 0.27(6) & 5.7 \\
$\nu_{45}$ & 38.4584(1)   & 0.021(2) & 0.86(3)  & 11.9  & $\nu_{109}$ & 18.5443(3) & 0.010(2) & 0.07(5) & 5.9 \\
$\nu_{46}$ & 43.2098(1)   & 0.021(2) & 0.26(3)  & 11.1  & $\nu_{110}$ & 24.1191(3) & 0.010(2) & 0.78(6) & 5.6 \\
$\nu_{47}$ & 45.7310(1)   & 0.021(2) & 0.02(3)  & 11.2  & $\nu_{111}$ & 26.9385(3) & 0.010(2) & 0.25(5) & 6.1 \\
$\nu_{48}$ & 51.0188(1)   & 0.021(2) & 0.72(3)  & 11.7  & $\nu_{112}$ & 27.8696(3) & 0.010(2) & 0.69(5) & 6.1 \\
$\nu_{49}$ & 21.6111(1)   & 0.020(2) & 0.29(3)  & 11.5  & $\nu_{113}$ & 31.8504(3) & 0.010(2) & 0.28(5) & 6.1 \\
$\nu_{50}$ & 39.8886(1)   & 0.020(2) & 0.17(3)  & 11.5  & $\nu_{114}$ & 33.7109(3) & 0.010(2) & 0.17(5) & 6.0 \\
$\nu_{51}$ & 43.2250(1)   & 0.020(2) & 0.03(3)  & 10.7  & $\nu_{115}$ & 34.3145(3) & 0.010(2) & 0.75(5) & 6.1 \\
$\nu_{52}$ & 32.1193(1)   & 0.019(2) & 0.12(3)  & 11.0  & $\nu_{116}$ & 39.8509(3) & 0.010(2) & 0.41(6) & 5.5 \\
$\nu_{53}$ & 33.7602(1)   & 0.019(2) & 0.37(3)  & 10.9  & $\nu_{117}$ & 40.0016(3) & 0.010(2) & 0.29(6) & 5.5 \\
$\nu_{54}$ & 39.3251(1)   & 0.019(2) & 0.29(3)  & 11.2  & $\nu_{118}$ & 41.7959(3) & 0.010(2) & 0.23(6) & 5.5 \\
$\nu_{55}$ & 48.6281(1)   & 0.019(2) & 0.57(3)  & 10.3  & $\nu_{119}$ & 43.1760(3) & 0.010(2) & 0.89(6) & 5.3 \\
$\nu_{56}$ & 1.9632(2)    & 0.018(2) & 0.67(4)  & 8.0   & $\nu_{120}$ & 43.8168(3) & 0.010(2) & 0.62(6) & 5.6 \\
$\nu_{57}$ & 46.0752(2)   & 0.018(2) & 0.79(3)  & 10.0  & $\nu_{121}$ & 47.7964(3) & 0.010(2) & 0.22(6) & 5.6 \\
$\nu_{58}$ & 49.9622(2)   & 0.018(2) & 0.66(3)  & 9.6   & $\nu_{122}$ & 52.6757(3) & 0.010(2) & 0.33(6) & 5.7 \\
$\nu_{59}$ & 2.5465(2)    & 0.017(2) & 0.54(4)  & 8.1   & $\nu_{123}$ & 53.6504(3) & 0.010(2) & 0.60(5) & 6.1 \\
$\nu_{60}$ & 32.0965(2)   & 0.017(2) & 0.24(3)  & 10.0  & $\nu_{124}$ & 19.1391(3) & 0.009(2) & 0.35(6) & 5.5 \\
$\nu_{61}$ & 33.1643(2)   & 0.017(2) & 0.69(3)  & 9.9   & $\nu_{125}$ & 25.5776(3) & 0.009(2) & 0.99(6) & 5.5 \\
$\nu_{62}$ & 33.7707(2)   & 0.017(2) & 0.26(3)  & 10.0  & $\nu_{126}$ & 33.4091(3) & 0.009(2) & 0.74(6) & 5.3 \\
$\nu_{63}$ & 36.7595(2)   & 0.017(2) & 0.08(3)  & 10.0  & $\nu_{127}$ & 34.2694(3) & 0.009(2) & 0.66(6) & 5.6 \\
$\nu_{64}$ & 39.8658(2)   & 0.017(2) & 0.25(3)  & 9.8   & $\nu_{128}$ & 47.3015(3) & 0.009(2) & 0.80(6) & 5.3 \\
\enddata

\tablecomments{Values in parentheses are the formal $1\sigma$ uncertainties in the last quoted digits. Frequencies are renumbered as $\nu_1,\nu_2,\ldots,\nu_{128}$ in descending order of amplitude. Phases are expressed as fractions of a pulsation cycle, modulo unity, relative to the adopted reference epoch. The table is arranged in two parallel blocks for compact presentation.}
\end{deluxetable*}

\section{\texorpdfstring{First order Companion light Tests}
{First order Companion light Tests}}
\label{app:missing_light}

To estimate the continuum contribution expected from a luminous
companion configuration, we used the same representative broadband bands
adopted in the SED analysis. The primary was fixed to the single star
solution, $T_{\rm eff,1}=8556~{\rm K}$ and
$R_1=2.789~R_\odot$. For a configuration containing one or more
main-sequence companion stars, the total continuum flux relative to the
primary was estimated at the effective wavelength of each band as
\begin{equation}
    \frac{F_{{\rm comp},\nu}}{F_{1,\nu}}
    =
    \frac{
    \displaystyle\sum_j R_j^2 B_\nu(T_{{\rm eff},j})
    }
    {R_1^2 B_\nu(T_{\rm eff,1})},
\end{equation}
where the sum contains one term for a single companion and two terms
for an unresolved inner binary. The total companion light fraction was
then calculated as
\begin{equation}
    \ell_{\rm comp}
    =
    \frac{F_{{\rm comp},\nu}}
    {F_{1,\nu}+F_{{\rm comp},\nu}},
\end{equation}
and the corresponding brightening relative to the primary-only
continuum is
\begin{equation}
    \Delta m
    =
    -2.5\log_{10}\!\left(
    1+\frac{F_{{\rm comp},\nu}}{F_{1,\nu}}
    \right).
\end{equation}

These calculations do not include full filter-throughput integration,
atmosphere-dependent line blanketing, or a simultaneous multi-component
SED fit. They are intended as first order continuum diagnostics of the
relative amount of companion light expected under different stellar
configurations.

\subsection{\texorpdfstring{Single Main-sequence Companion}
{Single Main-sequence Companion}}

For the single star hypothesis, we adopted representative
$T_{\rm eff,2}$ and $R_2$ values guided by empirical main-sequence
stellar-parameter relations
\citep{PecautMamajek2013,Eker2018}. The resulting light fractions are
listed in Table~\ref{tab:missing_light_single}.

\begin{deluxetable}{ccccccc}
\tablecaption{Estimated Continuum Light Contribution from a Single
Main-sequence Companion\label{tab:missing_light_single}}
\tablehead{
\colhead{$M_2$} &
\colhead{$T_{\rm eff,2}$} &
\colhead{$R_2$} &
\colhead{$\ell_{\rm comp}(G)$} &
\colhead{$\ell_{\rm comp}(K_s)$} &
\colhead{$\ell_{\rm comp}(W1)$} &
\colhead{$\Delta G$} \\
\colhead{($M_\odot$)} &
\colhead{(K)} &
\colhead{($R_\odot$)} &
\colhead{} &
\colhead{} &
\colhead{} &
\colhead{(mag)}
}
\startdata
1.70 &  7600 & 1.55 & 0.180 & 0.206 & 0.210 & $-0.216$ \\
1.80 &  8000 & 1.65 & 0.225 & 0.241 & 0.243 & $-0.277$ \\
1.86 &  8300 & 1.72 & 0.259 & 0.267 & 0.268 & $-0.326$ \\
2.00 &  9000 & 1.85 & 0.335 & 0.321 & 0.319 & $-0.443$ \\
2.30 &  9600 & 2.10 & 0.434 & 0.400 & 0.396 & $-0.617$ \\
3.00 & 12000 & 2.55 & 0.654 & 0.570 & 0.559 & $-1.152$ \\
\enddata
\end{deluxetable}

\subsection{\texorpdfstring{Unresolved Inner binary Companion}
{Unresolved Inner binary Companion}}

We extended the same calculation to a hierarchical configuration in
which the dynamical companion is an unresolved main-sequence inner binary.
The total inner binary mass was fixed to
$M_{\rm in}=M_a+M_b=1.80~M_\odot$, corresponding to the formal edge-on
dynamical lower limit. We define the inner binary mass ratio as
$q=M_b/M_a\leq1$. For each component, the luminosity and radius were
assigned using the empirical main-sequence mass--luminosity and
mass--radius relations of \citet{Eker2018}, and the effective temperature
was obtained from the Stefan--Boltzmann relation.

Table~\ref{tab:missing_light_binary} gives the predicted continuum
contribution for several representative mass ratios. An equal mass
$0.90+0.90~M_\odot$ inner binary contributes only
$\ell_{\rm comp}(G)\simeq0.038$, compared with
$\ell_{\rm comp}(G)=0.225$ for the representative single
$1.80~M_\odot$ companion in
Table~\ref{tab:missing_light_single}. Its contribution is larger at
infrared wavelengths, reaching approximately 0.083 in 2MASS $K_s$ and
0.091 in WISE W1, but remains substantially below that expected from the
single star configuration.

\begin{deluxetable}{cccccccc}
\tablecaption{Estimated Continuum Light Contribution from an
Unresolved Inner Binary with
$M_{\rm in}=1.80~M_\odot$\label{tab:missing_light_binary}}
\tablehead{
\colhead{$q$} &
\colhead{$M_a$} &
\colhead{$M_b$} &
\colhead{$M_a+M_b$} &
\colhead{$\ell_{\rm comp}(G)$} &
\colhead{$\ell_{\rm comp}(K_s)$} &
\colhead{$\ell_{\rm comp}(W1)$} &
\colhead{$\Delta G$} \\
\colhead{} &
\colhead{($M_\odot$)} &
\colhead{($M_\odot$)} &
\colhead{($M_\odot$)} &
\colhead{} &
\colhead{} &
\colhead{} &
\colhead{(mag)}
}
\startdata
1.0 & 0.900 & 0.900 & 1.800 & 0.038 & 0.083 & 0.091 & $-0.042$ \\
0.7 & 1.059 & 0.741 & 1.800 & 0.050 & 0.095 & 0.104 & $-0.056$ \\
0.5 & 1.200 & 0.600 & 1.800 & 0.076 & 0.124 & 0.132 & $-0.086$ \\
0.3 & 1.385 & 0.415 & 1.800 & 0.126 & 0.179 & 0.188 & $-0.146$ \\
\enddata
\end{deluxetable}

This calculation demonstrates that the broadband-light constraint
depends strongly on the architecture of the dynamical companion. A
near equal mass inner binary can supply the required total mass while
remaining substantially less prominent in the combined SED than a single
main-sequence star of the same total mass. The present SED therefore
provides stronger tension with the single star hypothesis than with a
near equal mass hierarchical configuration.

\bibliography{references_clean}
\bibliographystyle{aasjournalv7}

\end{document}